\documentclass[twocolumn,showpacs,preprintnumbers,amsmath,amssymb,prb,aps]{revtex4-1}
\usepackage{epsfig}
\usepackage{epstopdf}
\usepackage{multirow}
\usepackage{graphicx}
\usepackage{dcolumn}
\usepackage{bm}
\usepackage{textcomp}
\usepackage{float}
\usepackage{array}
\usepackage{csquotes}
\usepackage{subcaption}
\usepackage{booktabs}
\usepackage{amsmath}
\usepackage{hyperref}
\hypersetup{
    colorlinks=true,
    linkcolor=blue,
    filecolor=magenta,      
    urlcolor=blue,
    pdftitle={},
    }
    
\begin{document}
\title{Signatures of Hund's metal physics in single-layered 3$d$ transition metal oxide, $\mathrm{Sr_2CoO_4}$ }
\author{Shivani Bhardwaj$^{1,}$}
\altaffiliation{Electronic mail: sbhardwajiitm369@gmail.com}

\author{Sudhir K. Pandey$^{2,}$}
\altaffiliation{Electronic mail: sudhir@iitmandi.ac.in}
\affiliation{$^{1}$School of Physical Sciences, Indian Institute of Technology Mandi, Kamand - 175075, India}

\affiliation{$^{2}$School of Mechanical and Materials Engineering, Indian Institute of Technology Mandi, Kamand - 175075, India}

\date{\today} 

\begin{abstract}

With density functional theory plus dynamical mean-field theory, we study the influence of Hund's coupling on the nature of electronic correlations in $\mathrm{Sr_2CoO_4}$. Our results suggest strong signatures of Hund's metal physics in this compound. The Co 3$d$ states show large orbital differentiation in the degree of correlations and mass enhancement. The imaginary-time correlation functions suggest the presence of spin-orbital separation and large local charge fluctuations in the system. Breakdown of the Fermi-liquid picture is observed at the lowest calculated temperature for various strengths of Hund's coupling, suggesting the Fermi-liquid coherence scale lower than $\sim$100 K. Interestingly, a sudden emergence of a gapped state is noted for $e_g$ orbitals in its spectral density of states at $\sim$200 K in the vicinity of Fermi-level. Among the Co 3$d$ states 3$d_{z^2}$ and 3$d_{x^2-y^2}$ foster enlarged correlations. This study conclusively identifies $\mathrm{Sr_2CoO_4}$ as the first single-layered 3$d$ transition metal oxide to be classified as Hund's metal.
\vspace{0.2cm}

\end{abstract}

\maketitle


$\textit{Introduction}$. The theoretical classification of correlated materials has been traditionally based on the competing interactions between the hopping amplitude or bandwidth ($W$) of electrons and their on-site Coulomb repulsion ($U$). Typically, the interplay between these parameters categorizes the behavior of electrons into itinerant and localized. The localized electron systems are further cast into weakly($U$$<$$<$$W$), moderately ($U$$\sim$$W$) and strongly ($U$$>$$>$$W$) correlated systems based on the strength of $U$ \cite{1,2,3}.
The strongly correlated systems in particular have drawn immense attention as they exhibit rich variety of exotic phenomena ranging over unconventional superconductivity\cite{4}, heavy-fermionic behavior\cite{5}, non-Fermi liquid states\cite{6}, Mott-insulating behavior\cite{7,8,9}, Kondo physics\cite{kondo,kondo1,kondo2,kondo3}, which fail to be captured by conventional band theory\cite{band,band1, band2}.
These phenomena arise from strong electron-electron interactions, thus require theoretical treatment beyond band-theory based mean-field approaches. The understanding of electronic correlations in such systems is done by investigating the role of key interaction parameters that govern their electronic structure properties. Usually, $U$ and on-site exchange interaction or Hund's coupling ($J$) are considered as the critical parameters which essentially quantify the strength of intra-atomic electron-electron interactions in the system\cite{9}. \\
It is well recognized that strong correlations can arise due to large $U$ causing charge localization, famously known as Mott-insulating behavior\cite{m1,m2,m3, mott, mott1, mott2,mott3, mott4,mott5}. However, with the discovery of iron-based superconductors\cite{i1,i2,i3,i4,i5}, renewed interest in ruthenates\cite{r1,r2,r3} and several iron pnictides and chalcogenides\cite{i3,i5}, Hund's coulping was found to primarily drive strong correlations rather than $U$$\cite{h1,h2,h3}$.
This new class of materials is called Hund's metals, and an important feature of these materials includes $J$ induced interactions between spin, charge and orbital degrees of freedom\cite{o1,o2,o3,o4,o5}. 
These systems exhibit breakdown of Fermi-liquid behavior at low temperatures which is characterized by anomalous temperature dependencies such as absence of quasiparticle peaks at low temperatures, bad metal behavior in transport properties, etc\cite{r2}. In addition to this, the spin and orbital degrees of freedom in these systems undergo coherent to incoherent crossover at distinct  energy scales\cite{o1,o2,o3,o4,o5}.
Subsequently, this spin-orbit separation occurs due to different low-energy scales for spin and orbital excitaions which tends to cause orbital-selective Mott transitions observed in several of these multi-orbital systems such as $\mathrm{FeSe_{1-x}Te_x}$\cite{h1,o5} and $\mathrm{LaFeAsO}$\cite{i3}.

Nevertheless, few materials fall into the intermediate regime where both $U$ and $J$ are significant, leading to behaviors characteristic of both Mott and Hund’s physics for instance, Nickelates (e.g., $\mathrm{NdNiO_2}$, $\mathrm{LaNiO_3}$) \cite{ht1,ht2}, Iron Chalcogenides (e.g., $\text{FeSe}$, $\text{FeTe}$)\cite{i3,i5}.
However, numerous strongly correlated materials can be broadly categorized based on different origins of correlation effects ($U$ or $J$) into Mott-insulators or Hund's metals. 

In contrast to Mott physics, which is relatively well understood and has been extensively studied in systems for decades, the field of Hund's metals is still evolving as it constitutes a more complex, intermediate-coupling regime ($U$$\sim$$W$) in multiorbital systems.
For instance, Mott physics has been successful in explaining various phenomena such as high-temperature superconductivity in cuprates\cite{c1,c2}, quantum spin-liquid formation in several organics\cite{g1}, and kagome lattices\cite{k1}, apart from providing traditional insight into the insulating behaviour of several 3$d$ transition metal oxides, vanadium oxides\cite{m1},\& rare-earth and actinide oxides\cite{ht1,a1}. These Mott systems show dominant on-site Coulomb repulsion as large as $\sim$10 eV\cite{band1}. However, understanding of several other phenomena stemming from strong correlations such as orbital selectivity, strongly enhanced quasiparticle mass even in the absence of large $U$\cite{}, spin-orbital separation\cite{r3}, coherence-incoherence crossovers\cite{l1}, and non-Fermi liquid behavior\cite{h1} has been facilitated recently by understanding the critical role of Hund's coupling in these systems\cite{h1}.
 So far, the predominant role of Hund's coupling has been identified to explain unconventional superconductivity, non Fermi-liquid behavior in the metallic state of $\mathrm{Sr_2RuO_4}$\cite{r1,r2,r3} and $\mathrm{FeSe}$ \& $\mathrm{FeTe}$. Several other materials like Iron pnictides, Iron chalcogenides, chromium-based compounds such as $\mathrm{CrO_2}$$\cite{c1}$ and $\mathrm{Sr_2FeMoO_6}$$\cite{ms1}$ (double perovskite) were also described for the exhibition of spin-orbit separation and resulting orbital selective correlation effects based on the influence of $J$.
Notably, these systems demonstrate large strength of Hund's coupling, resulting in strong correlation effects in the presence of moderate $U$.
Considering our recent study on $\mathrm{Sr_2CoO_4}$, where the Hund's coupling strength is estimated to be as large as $\sim$ 1.16 eV with $U$ = 4.4 eV\cite{s1}, encourages us to examine the role of Hund's coupling in dictating the correlation effects in this system. Moreover, $\mathrm{Sr_2CoO_4}$ shares structure with well known 4$d$ transition metal oxide Hund's metal $\mathrm{Sr_2RuO_4}$ akin to $K_2NiO_4$ family. Nevertheless, $\mathrm{Sr_2CoO_4}$ (a quasi-two dimensional layered perovskite) is also reported to show anomalous behavior in its family of compounds as being the only member of $K_2NiO_4$ family to show ferromagnetic spin-response and metallic behavior\cite{fm}.
It should be noted that several magnetic spin states ($S$ = 3/2, $S$ = 1, $S$ = 1/2) have been reported by different theoretical studies, indicating complex nature of magnetism pressent in the system\cite{sr1,sr2,sr3,sr4,sr5,sr6}.  
The considerably large value of $J$ and also the intricate nature of magnetism in this system, indicates a plausible predominant role of $J$.

In this Letter, we examine $\mathrm{Sr_2CoO_4}$ to understand how $J$ affects the correlation effects in this system using the DFT+DMFT method, which effectively considers both band-like and atomic-like features on equal footing, making it a suitable approach to capture Hund's metal physics. We find strong signatures of Hundness present in the system, therefore putting it into the class of Hund's metals. This study finds $\mathrm{Sr_2CoO_4}$ to be the first 3$d$ transition metal layered perovskite of $\mathrm{K_2NiF_4}$ family to show the existence of Hund's metal physics.

\textit{Computational Details}. The electronic structure calculations of SCO within density functional theory + dynamical mean field theory (DFT+DMFT) framework is carried out with PBEsol exchange functional\cite{pbesol} where, augmented plane wave plus local orbitals (APW+lo) method is used to carry out the DFT calculations using WIEN2k code\cite{wien2k}. In the calculations, muffin-tin radii of 2.25, 1.82 and 1.57 a.u. are used for Sr, Co and O atoms, respectively. DFT+DMFT calculations are carried out with eDMFTF code \cite{edmft} wherein, the single-particle Green’s function is expanded in LAPW basis, and is fully charge self-consistently determined. WEIN2k performs the DFT calculation throughout DMFT iterations. The energy convergence criterion used is $\mathrm{10^{-4}}$ Ry/cell for 12 $\times$ 12 $\times$ 12 k-mesh size. Continuous-time QMC impurity solver is used with ‘exacty’ double-counting scheme\cite{exactd}. Subsequently, maximum-entropy method is used to calculate the spectral density for each Co 3$d$ orbital in real time/frequency domain.  Rotationally invariant type of Coulomb interaction paramterization (Full type) is used for the DFT+DMFT calculations. 
$\mathrm{Sr_{2}CoO_{4}}$ crystallizes in body-centered tetragonal (bct) and belongs to space group number $139$ (\textit{I4/mmm)}. Sr atoms are characterized by variable positions at $4e$ (0,0,z), and two inequivalent O atoms- O1 (planar) and O2 (apical) at $2b$ (0.5,0,0) and $4e$ (0,0,z), respectively. Co atoms occupy 2a (0,0,0) wyckoff positions. The calculations throughout have been performed with suppressed magnetic ordering to focus on the intrinsic role of correlations separately from magnetism in the system. The results are derived for $U$ = 4.4 and $J$ = 1.16 eV which are our main parameter choice (based on our previous study\cite{s1}). Experimental lattice parameters (a=b=3.79 $A^{0}$, c=12.48 $A^{0}$) \cite{sr1}with the relaxed positions of Sr (0.6423) and O2 (0.8445) are considered throughout calculations\cite{s1}.
 \begin{table*}[htbp!]
    \caption{Orbital-resolved mass-enhancement ($Z^{-1}$) of Co 3$d$ states for given $J$ values i.e. $J$ = 0, 0.9 and 1.16 eV}
    \begin{ruledtabular}
        \begin{tabular}{cccccccccc}
            $J$ (eV) &$d_{z^2}$ &$d_{x^2-y^2}$ & $d_{yz}/d_{xz}$ & $d_{xy}$   \\
            \hline
            0 &  1.20 & 1.30 & 2.50  & 2.47  \\
           
            0.9 & 2.14 & 2.67  & 2.46 & 2.44  \\
            
            1.16 & 3.68 & 4.69  & 3.14 & 3.0   \\
        \end{tabular}
    \end{ruledtabular}
\end{table*}

\textit{Results and discussion}. In order to study the role of Hund's coupling and resulting correlation effects, we first study the effect on mass enhancement of the Co 3$d$ states in the system.     
The mass enhancement ($\small m^*/m$) or the inverse quasiparticle weight ($Z^{-1}$) is calculated using the imaginary part of self-energy ($Im\Sigma(i\omega)$) in the matsubara frequency regime ($i\omega$) as below:
\begin{equation}
 m^*/m = Z^{-1} = 1 - \left. \frac{\partial \operatorname{Im} \Sigma(i\omega)}{\partial i\omega} \right|_{i\omega = 0}
\end{equation}

 Table I, displays the orbital-resolved $Z^{-1}$ of Co 3$d$ states for selected values of $J$ (0, 0.9, 1.16 eV) at $U$ = 4.4 eV.
It can be observed that when Hund's coupling is kept switched off, i.e., $J$ = 0, the $d_{z^2}$ and $d_{x^2-y^2}$ orbitals exhibit very small mass renormalisation, as inferred from their $Z$ values ($Z$$\sim$0.8), which do not deviate significantly from $Z$ = 1. In contrast, the $d_{xz/yz}$ and $d_{xy}$ orbitals undergo a nearly identical magnitude of $Z^{-1}$ ($m^*/m$ $\sim$2.5), resulting in their $Z$ values ($Z$ $\sim$0.4) being considerably lower than 1. This indicates significant mass renormalisation for the $d_{xz/yz}$ and $d_{xy}$ orbitals, even in the absence of Hund's coupling. Here, the estimates of the mass renormalisation factor clearly show a difference between the $e_g$ ($d_{z^2}$ and $d_{x^2-y^2}$) and $t_{2g}$ ($d_{xz/yz}$ and $d_{xy}$) orbitals at $J$ = 0. 

Further, with an increase in $J$ to 0.9 eV, there is a significant rise in $m^*/m$ for the $e_g$ orbitals ($d_{z^2}$ with $m^*/m$ = 2.14 and $d_{x^2-y^2}$ with $m^*/m$ = 2.67), while the $t_{2g}$ orbitals show a small decrease in their $m^*/m$ values for both the $d_{xz/yz}$ and $d_{xy}$ orbitals.
The results indicate an increase in orbital-selective Hund's induced correlations for the $e_g$ states compared to the $t_{2g}$ states, based on the mass enhancement range, which exceeds 2. 
However, with  further increase in the strength of $J$ to 1.16 eV, all the orbitals exhibit an enlarged $m^*/m$, with the $d_{x^2-y^2}$ state experiencing the largest $m^*/m$ ($\sim$4.7), followed by $d_{z^2}$ ($\sim$3.7), and the least enhancement shown by $d_{xy}$ (2.74). Note here that the orbitals within the $e_g$ and $t_{2g}$ states no longer exhibit similar magnitudes of $m^*/m$, resulting in distinct mass renormalisation for each orbital. This observation marks a clear differentiation of orbitals or a multi-orbital character, where the orbitals are affected differently by the Hund's coupling strength. This decoupling of orbitals, influenced by the effect of $J$, is characteristic of Hundness. Additionally, the range of $m^*/m$ observed in this system is consistent with previous estimates reported in Hund's metals\cite{i5,tom}.

Hund's metals often exhibit large charge fluctuations even while showing considerably significant $m^*/m$ or correlations\cite{tom,sig}. This typically happens close to the breakdown of the Fermi-liquid regime in these systems.
Therefore, we next study the results of local charge susceptibility and charge fluctuations in the system across an extensive temperature range (110-6000 K) to understand the degree of localisation of charges in $\mathrm{Sr_2CoO_4}$. 

Fig. 1 shows the temperature dependence of static local charge susceptibility ($\chi_{\text{charge}}$) given by-
\begin{equation}
  \chi_{\text{charge}}=\int_0^\beta \langle N_d(\tau) N_d(0) \rangle\,d\tau -\beta\langle N_d \rangle^2
\end{equation}
 and local charge fluctuations ($<$$\Delta$$N^2$$>$)
 \begin{equation}
   <\Delta N^2> = <\Delta N_d^2> - <\Delta N_d >^2
 \end{equation}  where $N_d$ is the total Co $3d$ occupancy, $\beta$ is inverse temperature. The total occupancy of the Co 3$d$ states over the studied range varies negligibly between 6.34 and 6.38. The results depict the weak temperature dependence of $<$$\Delta$$N^2$$>$ over the temperature range, changing by only $\sim$6\% over the studied range. We note similar behaviour in temperature-dependent $\chi_{\text{charge}}$ values, with no signatures of discontinuity or abrupt transition with temperature. The weak temperature dependence and considerably large estimates of $<$$\Delta$$N^2$$>$ ($\sim$0.72) and $\chi_{\text{charge}}$ again signify the presence of Hund's metal character\cite{h1,tom,sig}.


\begin{figure}[htbp!]
    \centering
    \includegraphics[width=6cm, height=6cm]{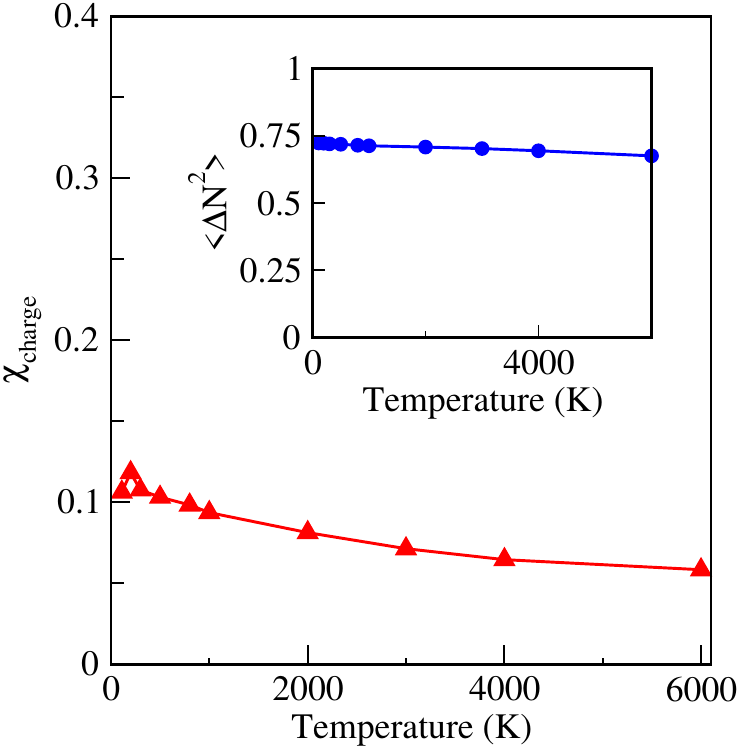}
    \caption{ \small Temperature dependence of static local charge-susceptibility ($\chi_{\text{charge}}$ in the unit of 1/eV ); Inset: Temperature dependence of local charge fluctuations $<$$\Delta$$N^2$$>$.}

    \label{fig:}
\end{figure}
\begin{figure}[htbp!]
    \centering
    \includegraphics[width=6cm, height=8cm]{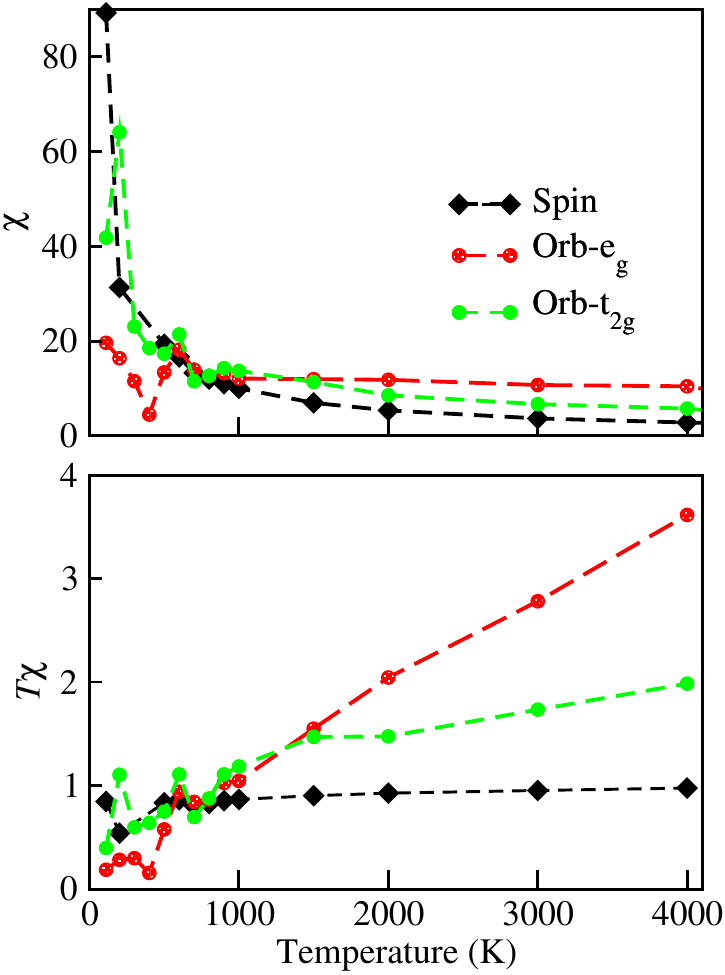}
    \caption{\small Temperature dependence of static local spin susceptibility ($\chi_{\text{spin}}$ in the unit of $\mu_B^{2}$/eV along with static local orbital susceptibility ($\chi_{\text{orb}}$ in the units of 1/eV for $e_g$ and $t_{2g}$ states }.

    \label{fig:}
\end{figure}
Furthermore, the system is examined for the presence of coherence-incoherence crossover, which is a common feature of Hund's metals. An incoherent to coherent crossover occurs at a characteristic low temperature where the system goes from an incoherent state at high temperatures to a coherent Fermi-liquid state at low temperatures. This is also indicated by the Curie-like behavior of magnetic spin susceptibility at high temperatures and crossing over to Pauli-like susceptibility below the characteristic coherent temperature\cite{r1,96,98,var,tom,sig}.

\begin{figure*}[htbp!]
    \centering
    \includegraphics[width=14cm, height=5.0cm]{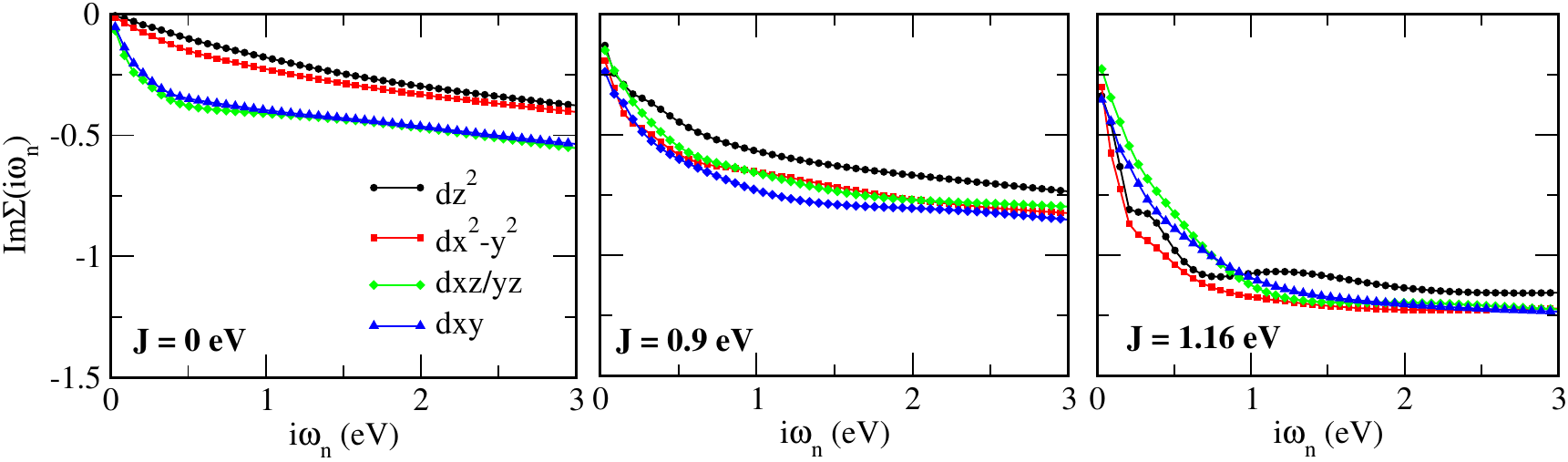}
    \caption{\small Orbital-resolved imaginary part of self-energy ($Im\Sigma(i\omega_n)$) for selected $J$s i.e. $J$ = 0, 0.9 \& 1.16 eV at $T$ = 110 K  }.

    \label{fig:}
\end{figure*}

We now study the imaginary-time correlation functions as static local spin and orbital susceptibilities (of both $e_g$ and $t_{2g}$ orbitals) for $\mathrm{Sr_2CoO_4}$ over the studied temperature range.
In Fig. 2, static local spin ( $\chi_{\text{spin}}$) and orbital ($\chi_{\text{orb}}$) susceptibilities are considered and given by 
\begin{equation}
  \chi_{\text{spin}}=\int_0^\beta \langle S^z(\tau) S^z(0) \rangle \, d\tau
\end{equation}
 and 
 \begin{equation}
   \chi_{\text{orb}}=\int_0^\beta \langle \Delta N_{\text{orb}}(\tau) \Delta N_{\text{orb}}(0) \rangle \, d\tau - \beta \langle \Delta N_{\text{orb}} \rangle^2
 \end{equation}. Here, $S_z$ is the total spin magnetic moment of Co $3d$ orbitals, $\Delta N_{\text{orb}}$ is the orbital occupancy difference per orbital ( for $e_g$ orbital $\Delta N_{\text{orb}}$ = $N_{z^2}$-$N_{x^2-y^2}$; for $t_{2g}$ orbital $\Delta N_{\text{orb}}$ = $N_{xz/yz}$-$N_{xy}$ ).
The $\chi_{\text{spin}}$ is observed to show Curie-like behavior above $T$ = 1000 K. Also, the $\chi_{\text{spin}}$ curve does not achieve saturation even reaching to the lowest temperature studied (110 K). 

Thus, the onset of screening of spin degrees of freedom could be regarded to start at $T^{o}$ = 1000 K (where $T^o$ is the onset of screening temperature) and completion of screening is expected to take place at $T$$<$100 K.
Whereas, $\chi_{\text{orb}}$ for $e_g$ and $t_{2g}$ orbitals show distinct trends across the studied temperature range. For instance, $\chi_{\text{orb}}$ of $t_{2g}$ states reveal a Curie-like behavior at extremely high temperatures, $T$ $>$ 2000 K, suggesting the onset screening temperature to be $T^{o}$$\sim$2000 K. On the other hand, $e_g$ states fail to show Curie behavior even till the highest temperature studied, indicating a much higher onset screening temperature $T^{o}$ $>$ 4000 K. This separation in the onset of screening temperatures of $\chi_{\text{spin}}$ and  $\chi_{\text{orb}}$ for both the orbitals i.e. $e_g$ and $t_{2g}$ show the evidence of spin-orbital separation in the system. It is essential to note that this separation is more pronounced in the case of $e_g$ states than $t_{2g}$ states. Moreover, the temperature dependent curves of $\chi_{\text{orb}}$ for both the orbitals do not saturate even at the lowest calculated temperature (110 K) suggesting that the onset of screening started at very high temperatures for both the orbitals ($T^o$$>$4000 K for $e_g$ and $T^o$$\sim$2000 K for $t_{2g}$) does not undergo complete screening at the lowest temperature considered in this study. 
Based on its temperature dependence curve, saturation in $\chi_{\text{spin}}$ and $\chi_{\text{orb}}$ is expected to be achieved below 110 K temperature, which indicates an even smaller Fermi-liquid coherence temperature scale.
  The large spin-orbit separation in the onset of screening arising in the $e_g$ sector than $t_{2g}$ can be a signature of its more correlated nature.

The above results suggest strong correlation effects resulting from the influence of strong Hund's coupling. To look into this more, Fig. 3 shows the imaginary part of self-energy ($Im\Sigma(i\omega_n)$; $i\omega$ is the imaginary frequency or Matsubara frequency) for the chosen $J$ values at $T$ = 110 K. The values of $Im\Sigma(i\omega_n)$ were obtained and found to reveal a complete breakdown of Fermi-liquid behaviour when fitted to a linear function in the low-frequency regions for each $J$, i.e., $F(i\omega_n)$ = C$i\omega_n$$^\alpha$, where $\alpha$ falls in the range 0.3-0.4 for all the orbitals at finite $J$ values ($J$ = 0.9 eV and 1.16 eV). At $J$ = 0, the $Im\Sigma(i\omega_n)$ of the $d_{xz/yz}$ and $d_{xy}$ orbitals show a linear behavior over an appreciable range of frequencies, while the $d_{z^2}$ and $d_{x^2-y^2}$ orbitals only show this straight-line pattern in a small low-frequency range before they start to change at frequencies above $i\omega_n > 1$ eV. Additionally, this separation in how the $e_g$ and $t_{2g}$ orbitals behave at $J$ = 0 changes when there is a finite Hund's exchange, leading to differences between the orbitals. Interestingly, the $d_{z^2}$ and $d_{x^2-y^2}$ orbitals are more affected by $J$, which is evident from the non monotonous behavior of their curves in $i\omega_n$$<$ 1 eV region.
Furthermore, the appearance of non-Fermi-liquid behaviour even at the low calculated temperature here ($T$ = 110 K) in each case ($J$ = 0.9 and 1.16 eV) is suggestive of the existence of a lower Fermi-liquid coherence scale as noted from the low temperature behavior of $\chi_{\text{spin}}$ and $\chi_{\text{orb}}$ as noted before. 
It is also noted that violation of Fermi-liquid behaviour is generally linked to $J$ in Hund's metals and is often attributed to $J$-induced spin and orbital fluctuations\cite{non,ruth,24}.  

\begin{figure*}[htbp!]
    \centering
    \includegraphics[width=11cm, height=6.0cm]{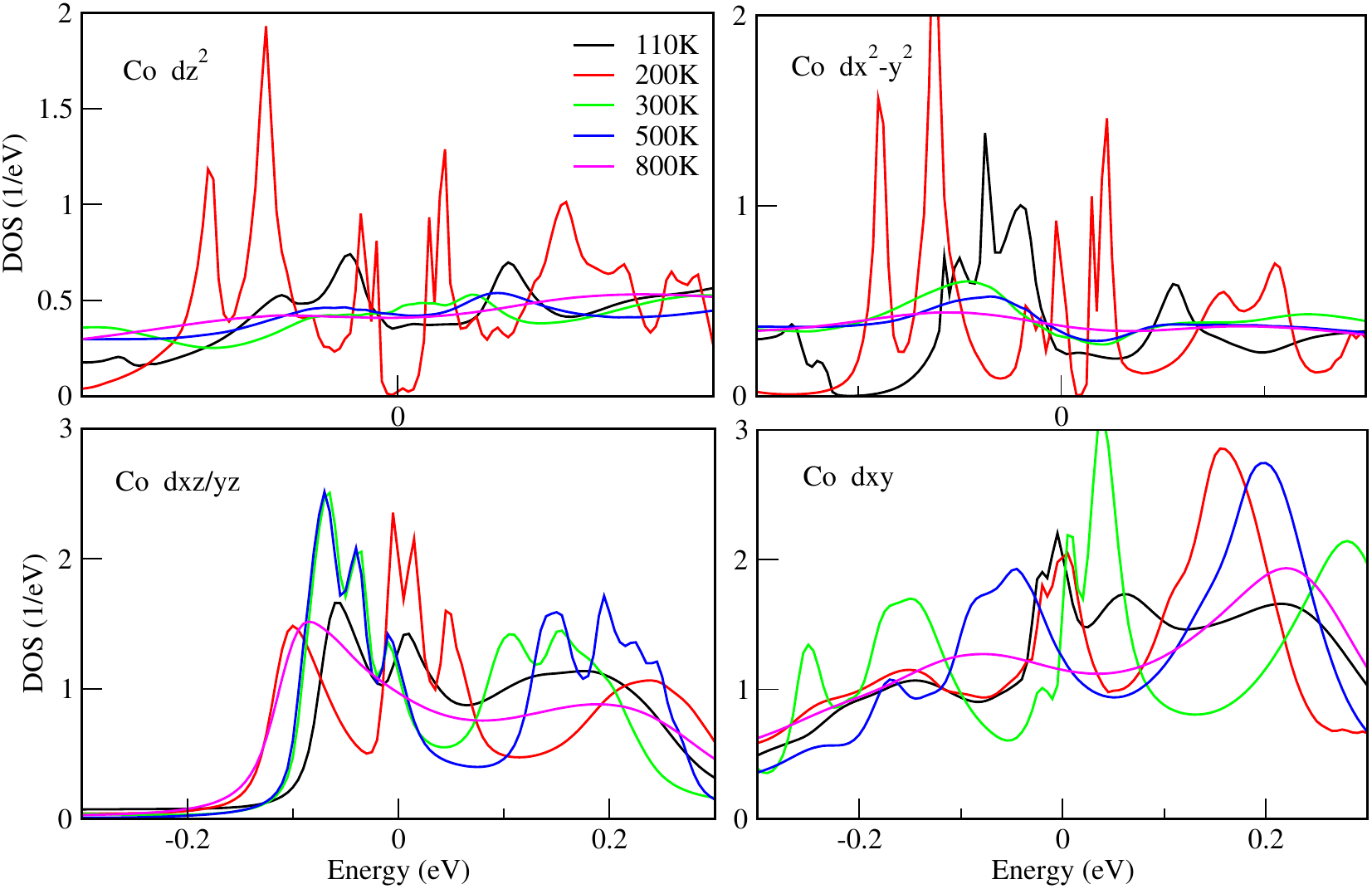}
    \caption{\small Orbital-resolved temperature dependent evolution of DOS of the Co 3$d$ states in close proximity of Fermi-level.}.

    \label{fig:}
\end{figure*}




Besides, we find interesting evolution of spectral density of $d_{z^2}$ orbital in close proximity of Fermi-level ($E_f$) which additionally proves its more correlated nature among all the Co 3$d$ orbitals in this systems.
In Fig. 4. orbital-resolved results from the temperature-dependent evolution of the spectral density of states (DOS) in the proximity of $E_f$ are plotted. As evident from the figure the $d_z^2$ orbital displays sudden emergence of a small band-gap at $T$ = 200 K. However, while increasing temperature, incoherent weight begins to appear at $E_f$ for all $T$ $>$ 200 K. Interestingly, the opening of this small band-gap only arises for Hund's coupling strength of $J$ = 1.16 eV.a
It is also essential to note that a similar evolution in density of states takes place in $d_{x^2-y^2}$ orbital also except that the appearance of the gap in the DOS appears slightly from the $E_f$ in the unoccupied region.
On the contrary, $d_{xz/yz}$ and $d_{xy}$ orbitals show the presence of coherent weight and quasiparticle peak in the vicinity of $E_f$ at all temperatures $T$$<$800 K and with increasing temperature the coherent weight begins to migrate to higher energies away from the $E_f$.
The appearance of a gapped state in $e_g$ for $J$ = 1.16 eV  suggests abnormal temperature evolution of $d_{z^2}$ and $d_{x^2-y^2}$ orbitals driven by strong Hund's coupling. Conclusively, visibly large sensitivity of $e_g$ on $J$ distinguishes it for fostering Hund's induced gapped state and consequently large correlation effects. The atypical temperature dependence observed the DOS of $e_g$ states could be further investigated in relation to other electronic structure and transport properties to understand the combined $J$ and temperature effects in $\mathrm{Sr_2CoO_4}$.

$\textit{Conclusion}$. We demonstrate the presence of characteristic signatures of Hundness in $\mathrm{Sr_2CoO_4}$, which include: (i)
observation of intermediate mass enhancement for all Co 3$d$ states Where, $d_{z^2}$ and $d_{x^2-y^2}$ ($e_g$) orbitals undergo more drastic mass enhancement than $d_{xz/yz}$ and $d_{xy}$ ($t_{2g}$) orbitals,  (ii)
presence of high local charge susceptibility and charge fluctuations showing characteristic itinerant nature in spite of considerable mass enhancement,
(iii) breakdown of Fermi-liquid behaviour even at the lowest calculated temperature ($\sim$100 K) and
(iv) large spin-orbit separation derived from imaginary-time correlation functions and indications of lower Fermi-liquid coherence scale ($T$$<$100 K).
An interesting gapped state emerges in the $d_{z^2}$ and $d_{x^2-y^2}$ orbitals at a temperature scale of 200 K.
Conclusively, the $e_{g}$ orbitals show pronounced correlation effects as compared to $t_{2g}$ orbitals, suggesting existence of orbital selective Hund's driven correlation effects in the system.
Our study thereby classifies $\mathrm{Sr_2CoO_4}$ as Hund's metal. \\

$\textit{Acknowledgement}$. We acknowledge the computational support provided
by the High-Performance Computing (HPC) PARAM Himalaya at the Indian Institute of Technology Mandi.

\end{document}